\documentclass[preprint,nofootinbib,preprintnumbers]{revtex4}
\usepackage{amssymb}
\usepackage{graphicx}

\preprint{\hspace{13.45cm} NT@UW-04-029}

\begin{document}

\title{Universality of the EMC Effect}
\author{Jiunn-Wei Chen}
\email{jwc@phys.ntu.edu.tw}
\affiliation{Department of Physics and National Center for Theoretical Sciences at
Taipei, National Taiwan University, Taipei 10617, Taiwan, \\
and Lawrence-Berkeley Laboratory, Berkeley, CA 94720, U.S.A.}
\author{William Detmold}
\email{wdetmold@phys.washington.edu}
\affiliation{Department of Physics, University of Washington, Seattle, WA 98195-1560,
U.S.A.}
\date{\today }

\begin{abstract}
Using effective field theory, we investigate nuclear modification of nucleon
parton distributions (for example, the EMC effect). We show that the
universality of the shape distortion in nuclear parton distributions (the
factorisation of the Bjorken $x$ and atomic number ($A$) dependence) is
model independent and emerges naturally in effective field theory. We then
extend our analysis to study the analogous nuclear modifications in isospin
and spin dependent parton distributions and generalised parton distributions.
\end{abstract}

\maketitle

In 1983, the European Muon Collaboration (EMC) reported measurements \cite%
{Aubert:1983xm} of the ratio $R_{Fe}(x)$ of $F_{2}(x)$ structure functions
in iron and deuterium in deep inelastic scattering (DIS). In the parton
model, these structure functions are defined as $F_{2}^{A}(x)=%
\sum_{i}Q_{i}^{2}x\left[ q_{i}^{A}(x)+\overline{q}_{i}^{A}(x)\right] $,
where $q_{i}^{A}(x)$[$\overline{q}_{i}^{A}(x)$] is the parton distribution
function (PDF) for quarks[anti-quarks] of flavor $i$ in a nucleus, $A$, and $%
Q_{i}$ is the electric charge of $q_{i}$. Bjorken $x$ is the longitudinal
momentum fraction of the struck parton in the infinite momentum frame with
respect to its parent nucleon. $R_{Fe}(x)$ is normalised to unity if the
nucleons in the nuclei are non-interacting. The main observation of EMC was
that $R_{Fe}(x)$ deviated from unity by up to 20\% over the range $%
0.05<x<0.65$ in which measurements were taken. The EMC result was unlooked
for and came as a surprise to many physicists at the time partly because the
typical binding energy per nucleon is so much smaller ($<$1\%) than the
nucleon mass and the energy transfer involved in a DIS process.

Over the past two decades, further experiments have been performed by many
groups (see Refs.~\cite%
{Arneodo:1992wf,Geesaman:1995yd,Piller:1999wx,Norton:2003cb} for recent
reviews) aiming to better understand the details of nuclear modifications of
hadron structure functions (here referred to generically as the EMC effect).
The EMC result has been confirmed and demonstrated in many other nuclei
ranging from helium to lead. Nuclear modification of structure functions has
also been studied in other situations such as proton-nucleus Drell-Yan
experiments and quarkonium production \cite{Peng:1999tm,Norton:2003cb}.
Overall, a very interesting picture has emerged; for an isoscalar nucleus of
atomic number $A$, the \emph{shape} of the deviation from unity of $%
R_{A}(x)=F_{2}^{A}(x)/AF_{2}^{N}(x)$ (here $F_{2}^{d}$ has been converted to
the isoscalar $F_{2}^{N}$ subject to a small model dependent error, and the
slight $Q^{2}$ dependence of $R_{A}(x)$ \cite{Norton:2003cb} is suppressed)
is universal \cite{Date:1984ve,Frankfurt:1988nt}, namely independent of $A$
within experimental error bars, while the \emph{magnitude} of the distortion
is empirically proportional to the number density of the nucleus, $\rho _{A}$
\cite{Gomez:1993ri}. Fits to the available data that support these features
will be presented elsewhere \cite{futurework}. These findings have inspired
many theoretical analyses seeking to understand the details of the EMC
effect in various approaches (see the reviews for summaries). Different
physical processes have been identified as the causes of the modifications
in different $x$ regions; \textit{e.g.}, nuclear shadowing at low $x$ and
Fermi motion at large $x$. A recent conclusion of detailed model studies was
that the EMC effect necessarily implies modification of the nucleon PDFs and
cannot be explained through traditional nuclear physics \cite{Smith:2003hu}.
However, little is known directly from QCD.

In this Letter, we employ effective field theory (EFT) to investigate the
EMC effect by studying nuclear matrix elements of the twist-two operators
which are related to parton distributions and structure functions via the
operator product expansion. We find that the universality of the shape
distortion of the EMC effect is a model independent result, arising from the
symmetries of QCD and the separation of the relevant scales. The $x$
dependence of $R_{A}(x)$ is governed by short distance physics, while the
overall magnitude (the $A$ dependence) of the EMC effect is governed by long
distance matrix elements calculable using traditional nuclear physics. We
then proceed to study analogous nuclear effects in isospin-odd, and
spin-dependent parton distributions and generalised parton distributions
(GPDs) \cite{GPD}. We also discuss aspects of extracting the shape of the
EMC effect from first principles using lattice QCD. This approach provides a
clear connection between the EMC effect and many other observations of
nuclear modification of hadron properties.

EFT is a model independent approach which only makes use of the symmetries
and scale separation of the system (for recent reviews see Ref.~\cite%
{EFTreviews}). This approach has been successfully applied to many low
energy processes in $A=1,2,3,4$ systems. Recently EFT has been applied
to the computation of hadronic matrix elements of twist-two operators
in the meson and single nucleon sectors \cite{AS,CJ,Detmold:2005pt}
and applied to chiral extrapolations of lattice calculations of
moments of parton distributions \cite{DMNRT}. The approach has also
been extended to analyse moments of generalised parton
distributions~\cite{Jq}, large $N_{C}$ relations among PDFs in
nucleons and the $\Delta $-isobar \cite{CJ}, and deeply virtual
Compton scattering in the nucleon \cite{DVCSpi} and deuteron systems
\cite{BS}. The method is readily generalised to the multi-nucleon
case. Although we will concentrate on quark bilinear twist-two
operators, the framework can be easily applied to gluonic operators
and thereby to nuclear effects in gluonic distributions which are
important in heavy ion collisions at RHIC and LHC and at a future
Electron-Ion Collider.

To described the EMC effect observed in $F_2$ data on isoscalar nuclei, we
consider the normalised, spin singlet, isoscalar twist-two operators, 
\begin{equation}
\mathcal{O}_{q}^{\mu _{0}\cdots \mu _{n}}=\overline{q}\gamma ^{(\mu
_{0}}iD^{\mu _{1}}\cdots iD^{\mu _{n})}q/\left( 2M^{n+1}\right) ,  \label{O}
\end{equation}
where $(...)$ indicates that enclosed indices have been symmetrised
and made traceless, $D^{\mu }=(\overrightarrow{D}^{\mu }-
\overleftarrow{D}^{\mu })/2$ is the covariant derivative and $M$ is
the nucleon \ mass. The matrix elements of $\mathcal{O}_{q}^{\mu
  _{0}\ldots \mu _{n}}$ in an unpolarised single nucleon state with
momentum $P$ can be parametrised as
\begin{equation}
\langle P|\mathcal{O}_{q}^{\mu _{0}\ldots \mu _{n}}|P\rangle =\langle
x^{n}\rangle _{q}\widetilde{v}^{\mu _{0}}\cdots \widetilde{v}^{\mu _{n}},
\end{equation}
where the nucleon velocity $\widetilde{v}^{\mu }=P^{\mu }/M$. It is well
known that the coefficients $\langle x^{n}\rangle _{q}$ correspond to
moments of the isoscalar combination of parton distribution functions, 
\begin{equation}
\left\langle x^{n}\right\rangle _{q}=\int_{-1}^{1}dx\,x^{n}q(x),
\end{equation}
where $q(x)$ is the isoscalar quark distribution and $q(-x)=-\overline{q}(x)$.

We first consider only nucleonic degrees of freedom ({\it i.e.},
assume that pions are integrated out of the EFT -- they will be
reintroduced below) and perform the standard matching procedure in
EFT, equating the quark level twist-two operators to the most general
combinations of hadronic operators of the same symmetries
\cite{AS,CJ,Detmold:2005pt,DMNRT}. The leading one- and two-body
hadronic operators in the matching are
\begin{equation}
\mathcal{O}_{q}^{\mu _{0}\ldots \mu _{n}}=\langle x^{n}\rangle _{q}v^{\mu
_{0}}\cdots v^{\mu _{n}}N^{\dagger }N\left[ 1+\alpha _{n}N^{\dagger }N\right]
+\cdots \,,  \label{eq:1}
\end{equation}%
where $v^{\mu }=\widetilde{v}^{\mu }+\mathcal{O}(1/M)$ is the velocity
of the nucleus. Operators involving additional derivatives are
suppressed by powers of $M$ in the EFT power-counting. In
Eq.~(\ref{eq:1}) we have only kept the SU(4) (spin and isospin)
singlet two-body operator $\alpha _{n}v^{\mu _{0}}\cdots v^{\mu
  _{n}}\left( N^{\dagger }N\right) ^{2}$ in the above equation. The
other independent two-body operator $\beta _{n}v^{\mu _{0}}\cdots
v^{\mu _{n}}\left( N^{\dagger }\mbox{\boldmath$\tau$}N\right) ^{2}$,
which is non-singlet in SU(4) [$\mbox{\boldmath$\tau$}$ is an isospin
matrix], is
neglected because $\beta _{n}/\alpha _{n}=O(1/N_{c}^{2})\simeq 0.1$ \cite{KS}%
, where $N_{c}$ is the number of colors. Furthermore, the matrix element of $%
\left( N^{\dagger }\mbox{\boldmath$\tau$}N\right) ^{2}$ for an
isoscalar state with atomic number $A$ is smaller than that of $\left(
  N^{\dagger }N\right) ^{2}$ by a factor $A$ \cite{futurework}.
Three- and higher- body operators also appear in Eq.~(\ref{eq:1});
numerical evidence from other EFT calculations indicates that these
contributions are generally much smaller than two-body ones
\cite{Kubodera}.

\begin{figure}[t]
\begin{center}\includegraphics[width=0.3\columnwidth]{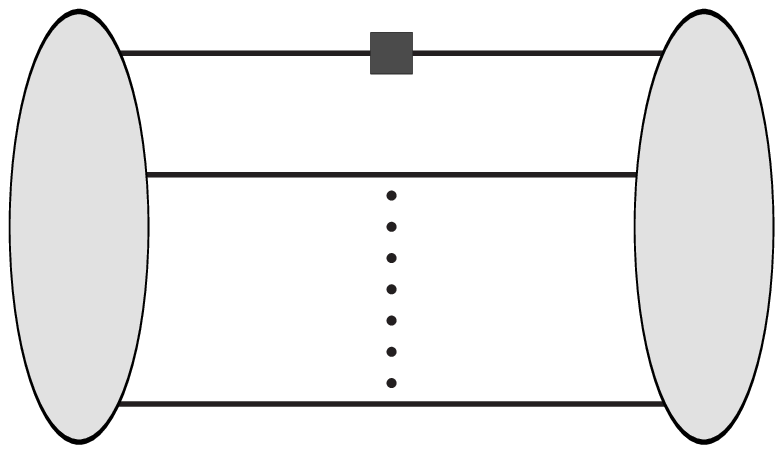} \hspace*{5mm} 
\includegraphics[width=0.3\columnwidth]{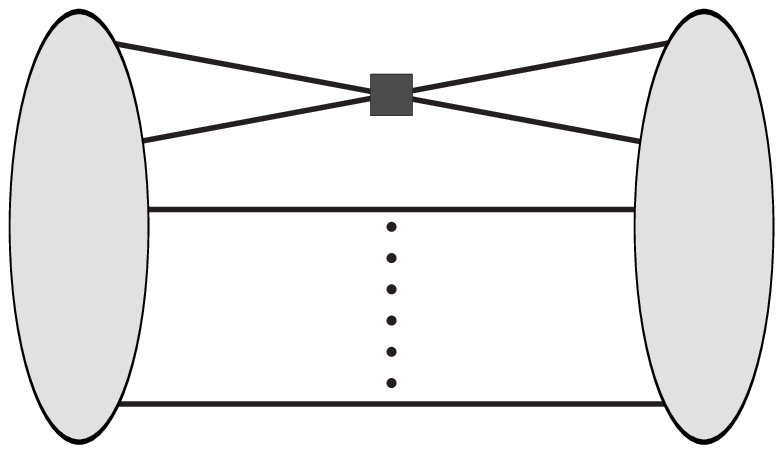}\\
(a)\hspace*{5.5cm}(b) \\
\centering \includegraphics[width=0.3\columnwidth]{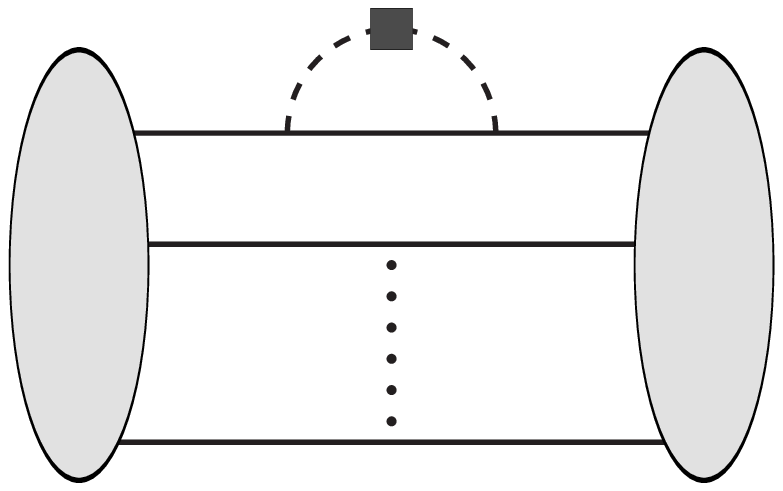} \hspace*{2mm} 
\includegraphics[width=0.3\columnwidth]{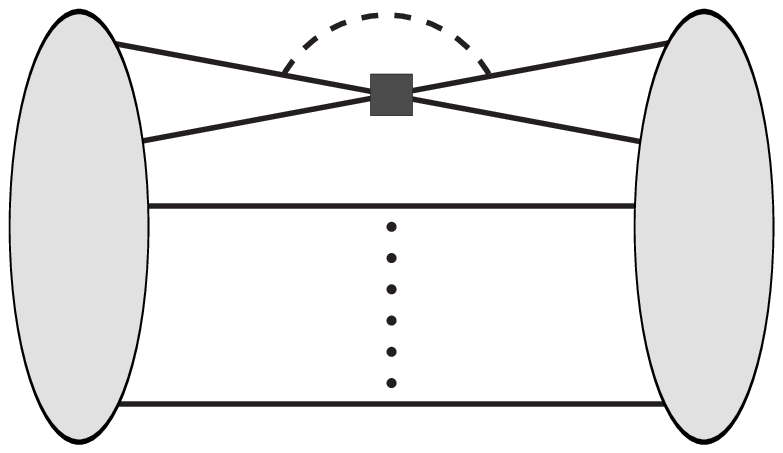} \hspace*{2mm} 
\includegraphics[width=0.3\columnwidth]{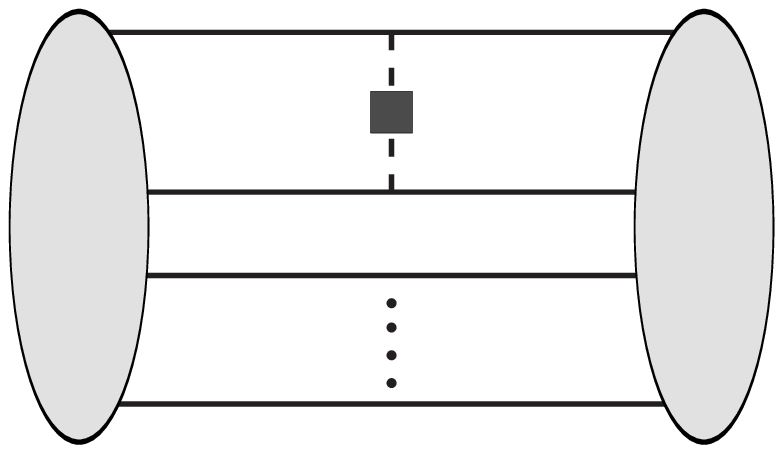}\\
(c)\hspace*{5.3cm}(d) \hspace*{5.3cm} (e) \vspace*{-2mm}
\end{center}
\caption{Contributions to nuclear matrix elements. The dark square
represents the various operators in Eq.~(\protect\ref{eq:1}) and the light
shaded ellipse corresponds to the nucleus, $A$. The dots in the lower part
of the diagram indicate the spectator nucleons.}
\label{fig:nucleon}
\end{figure}
Nuclear matrix elements of $\mathcal{O}_{q}^{\mu _{0}\ldots \mu _{n}}$ give
the moments of the isoscalar nuclear parton distributions, $q_{A}(x)$. The
leading order (LO) and the next-to-leading order (NLO) contributions to
these matrix elements are shown in Fig. 1(a) and 1(b), respectively. For an
unpolarized, isoscalar nucleus, 
\begin{eqnarray}
\langle x^{n}\rangle _{q|A} &\equiv &v^{\mu _{0}}\cdots v^{\mu _{n}}\langle
A|\mathcal{O}_{q}^{\mu _{0}\ldots \mu _{n}}|A\rangle  \label{eq:2} \\
&=&\langle x^{n}\rangle _{q}\left[ A+\langle A|\alpha _{n}(N^{\dagger
}N)^{2}|A\rangle \right] ,  \nonumber
\end{eqnarray}%
where we have used $\langle A|N^{\dagger }N|A\rangle =A$. Notice that if
there were no EMC effect, the $\alpha _{n}$ would vanish for all $n$. Also $%
\alpha _{0}=0$ because of charge conservation. Asymptotic relations \cite%
{West:1985kg} and analysis of experimental data \cite%
{Arneodo:1992wf,Rinat:2005qk} suggests that $\alpha _{1}\simeq 0$, implying
that quarks carry very similar fractions of a nucleon's and a nucleus's
momentum though no symmetry guarantees this. From Eq.~(\ref{eq:2}) we see
that the ratio 
\begin{equation}
\frac{{\displaystyle{\frac{\langle x^{n}\rangle _{q|A}}{A\langle
x^{n}\rangle _{q}}}}-1}{{\displaystyle{\frac{\langle x^{m}\rangle _{q|A}}{%
A\langle x^{m}\rangle _{q}}}}-1}=\frac{\alpha _{n}}{\alpha _{m}}
\label{eq:3}
\end{equation}%
is independent of $A$ which has powerful consequences. In all generality,
the isoscalar nuclear quark distribution can be written as 
\begin{equation}
q_{A}(x)=A\left[ q(x)+\widetilde{g}(x,A)\right] .  \label{eq:4}
\end{equation}%
Taking moments of Eq.~(\ref{eq:4}), Eq.~(\ref{eq:3}) then demands that the $%
x $ dependence and $A$ dependence of $\widetilde{g}$ factorise, 
\begin{equation}
\widetilde{g}(x,A)=g(x)\mathcal{G}(A),
\end{equation}%
with 
\begin{equation}
\mathcal{G}(A)=\langle A|(N^{\dagger }N)^{2}|A\rangle /A\Lambda _{0}^{3},
\label{G}
\end{equation}%
and $g(x)$ satisfying 
\begin{equation}
\alpha _{n}=\frac{1}{\Lambda _{0}^{3}\langle x^{n}\rangle _{q}}%
\int_{-A}^{A}dx\,x^{n}g(x)\,.  \label{an}
\end{equation}%
$\Lambda _{0}$ is an arbitrary dimensionful parameter and will be chosen as $%
\Lambda _{0}=1$~fm$^{-1}$. Crossing symmetry dictates that the even and odd $%
\alpha _{n}$ separately determine the nuclear modifications of valence and
total quark distributions. These results apply to any isoscalar combination
of parton distributions including $F_{2}(x)$ for isoscalar nuclei. Thus our
result implies that 
\begin{equation}
R_{A}(x)=\frac{F_{2}^{A}(x)}{AF_{2}^{N}(x)}=1+g_{F_{2}}(x)\mathcal{G}(A),
\label{Ra}
\end{equation}%
which says that the EMC effect (the deviation of $R_{A}(x)$ from unity) has
an universal shape described by $g_{F_{2}}(x)$ while the magnitude of the
deviation, $\mathcal{G}(A)$, only depends on $A$.

The above analysis gives a simple explanation of the observed universal
shape of the EMC effect, or equivalently, the factorisation of $\widetilde{g}%
(x,A)$. The key to establishing this factorisation is that other sources of
nuclear modification contributing to the right-hand side of Eq.~(\ref{eq:1})
must be suppressed (higher order in the EFT) such that the $A$ independence
of Eq.~(\ref{eq:3}) can be established. We stress that the factorisation
persists when pions are included in our analysis. In Fig.~\ref{fig:nucleon},
examples of the leading pionic contributions are shown. The various
single-nucleon diagrams, such as Fig.~\ref{fig:nucleon}(c), simply renormalise
the nucleon moments, $\langle x^{n}\rangle _{q}$, without contributing to
the EMC effect. Two- and more- nucleon diagrams such as those in Fig.~\ref%
{fig:nucleon}(d) and \ref{fig:nucleon}(e) contribute to the EMC
effect, but only at N$^{3}$LO and higher (see Ref.~\cite{futurework}
for explicit calculations) since the pions must be radiation pions
rather than potential pions \cite{KSW98} (we are free to choose the
twist-two indices to be $\mu _{i}=0$ for all $i$). Other contributions
that could upset the factorization include a two-body operator which
is similar to that in Eq.~(\ref{eq:1}) but
with two more derivatives. However this operator also contributes at N$^{3}$%
LO. Consequently, the universality of the EMC effect is preserved to good
accuracy. For large $x$ it is clear that the
factorisation must break down (simply consider the region $x>2$ in which
only three- and higher- body operators contribute) though the structure
function is very small in this region anyway. We stress that the
factorisation is a model independent result and is just a consequence of
scale separation and the SU(4) spin-isospin symmetry from large $N_{c}$.

It is clear from Eq.~(\ref{G}) that $\mathcal{G}(A)$ is governed by long
distance physics which can be computed using a traditional, non-relativistic
nuclear physics approach. It is interesting to note that the mean field
scaling of $\mathcal{G}(A)\sim \rho _{A}\sim \log (A)$ describes the
empirical $A$ dependence of $R_{A}(x)$ well \cite{Gomez:1993ri,futurework},
even though the mean field approximation is not justified for nuclei where
the two-particle S-wave scattering lengths are much larger than the mean
distance between nucleons.

Information on the shape distortion function $g(x)$ is encoded in the short
distance parameters $\alpha _{n}$ associated with the strength of the
two-body currents. One can either fix the $\alpha _{n}$ from experimental
data (to determine all $\alpha _{n}$, data on $F_{2}^{A}(x)$ and $%
F_{3}^{A}(x)$ are required) or calculate $\langle NN|\mathcal{O}_{q}^{\mu
_{0}\ldots \mu _{n}}|NN\rangle $ in two nucleon systems to extract them. The
latter approach, however, is intrinsically non-perturbative and thus
requires lattice QCD. Recent analysis \cite{Detmold:2004kw} suggest that one
can use background fields coupling to twist-two operators to extract
information on the $\alpha _{n}$ from the finite volume scaling of two
particle energy levels \cite%
{Luscher:1986pf,Beane:2003da,Detmold:2004qn,futurework}. At present, only
the first few coefficients could be extracted on the lattice because of
problems with operator mixing \cite{Gockeler:1996mu}. However, even a
calculation of these would be significant since it would be addressing
nuclear modification of hadron structure from first principles. Lattice
calculations can also be used to investigate the large $N_{c}$ prediction of 
$\beta_{n}/\alpha_{n} = O(1/N_{c}^{2})$ discussed below Eq.~(\ref{eq:1}).

Given the success of the EFT approach in explaining aspects of the
isoscalar, helicity averaged EMC effect, we shall now proceed to study the
isospin and spin dependent cases. For the isovector operators 
\begin{equation}
\mathcal{O}_{q,3}^{\mu _{0}\cdots \mu _{n}}=\overline{q}\gamma ^{(\mu
_{0}}iD^{\mu _{1}}\cdots iD^{\mu _{n})}\mbox{\boldmath$\tau$}_{3}q/\left(
2M^{n+1}\right) ,
\end{equation}%
EFT operator matching leads to%
\begin{equation}
\mathcal{O}_{q,3}^{\mu _{0}\ldots \mu _{n}}=\langle x^{n}\rangle
_{q,3}v^{\mu _{0}}\cdots v^{\mu _{n}}N^{\dagger }\mbox{\boldmath$\tau$}_{3}N%
\left[ 1+\alpha _{3,n}N^{\dagger }N\right] +\cdots \,,
\label{eq:isovec_operators}
\end{equation}%
where the ellipsis includes the higher order nucleonic and pionic operators.
Analysis similar to the isoscalar case implies 
\begin{equation}
R_{A}^{(3)}(x)\equiv \frac{q_{3|A}(x)}{(Z-N)q_{3}(x)}=1+g_{3}(x)\mathcal{\ G}%
_{3}(A)\,,
\end{equation}%
where $q_{3}(x)=u(x)-d(x)$ and $Z$($N$) is the proton (neutron) number. To
test this factorisation, one can either consider the difference between $%
F_{2}$'s in $(Z,N)=(n+m,n)$ and $(n,n+m)$ mirror nuclei \cite{Saito:2000fx}
and compare it with $F_{2}^{p}-F_{2}^{n}$, or disentangle $u_{A}(x)$ and $%
d_{A}(x)$ with the proposed neutrino-nucleus experiment, MINER$\nu $A \cite%
{Minerva}.

To generalise our analysis to spin dependent operators is also
straightforward. The operators related to moments of quark helicity
and transversity distributions are
\begin{eqnarray}
\mathcal{O}_{\Delta q,\alpha }^{\mu _{0}\cdots \mu _{n}} &=&\overline{q}%
\gamma ^{(\mu _{0}}\gamma ^{5}iD^{\mu _{1}}\cdots iD^{\mu _{n})}\tau
_{\alpha }q/\left( 2M^{n+1}\right) ,  \label{moreOs} \\
\mathcal{O}_{\delta q,\alpha }^{\rho \mu _{0}\cdots \mu _{n}} &=&\overline{q}%
\sigma ^{\rho (\mu _{0}}\gamma ^{5}iD^{\mu _{1}}\cdots iD^{\mu _{n})}\tau
_{\alpha }q/\left( 2M^{n+2}\right) \,.\;\;\;\;\;  \nonumber
\end{eqnarray}%
In the isoscalar case ($\tau _{0}=\mbox{\boldmath$1$}$), the matching yields 
\begin{equation}
\mathcal{O}_{\Delta q,0}^{\mu _{0}\cdots \mu _{n}}=2\langle x^{n}\rangle
_{\Delta q}N^{\dagger }S^{(\mu _{0}}v^{\mu _{1}}\cdots v^{\mu _{n})}N\left[
1+\gamma _{n,0}N^{\dagger }N\right] +\cdots ,  \nonumber
\end{equation}%
and similarly for the other operators (the spin operator $S^{\mu }=(0,%
\mbox{\boldmath$\sigma$}/2)$ where $\mbox{\boldmath$\sigma$}$ are Pauli spin
matrices). Again, we have similar factorisation; \textit{e.g.}, 
\begin{equation}
R_{A}^{\Delta }(x)\equiv \frac{\Delta q^{A}(x)}{\Delta N\,\Delta q(x)}%
=1+g_{\Delta }(x)\mathcal{G}_{\Delta }(A),
\end{equation}%
where $\Delta N$ is the differences between the number of nucleons with
positive and negative spin projections in the longitudinal direction.
Similarly, $\mathcal{G}_{\Delta }(A)=\langle A|N^{\dagger }NN^{\dagger }%
\mbox{\boldmath$\sigma$}_{3}N|A\rangle /\Delta N\Lambda _{0}^{3}$. Whilst
there is data on longitudinal asymmetries in light nuclei from which the $%
g_{1}^{A}(x)$ structure functions can be extracted, disentangling
nuclear effects in the unpolarised and polarised structure functions
will be difficult. Currently, polarised heavy nuclei targets, in which
modifications would be larger, are not available. Recent model
calculations \cite{polarisedmodels} find nuclear effects in the
polarised structure function $g_1^A(x)$ to be significant. Nothing is
known experimentally about the transversity structure function even in
the proton but analogous nuclear modifications can be derived.

It is also possible to study nuclear effects in GPDs by computing
off-forward matrix elements of twist-two operators \cite{futurework}.
In the quark contribution to nuclear spin, $J_{qA}$, for example, in
addition to extending the operators matched to
$\mathcal{O}_{q}^{\mu_{0}\mu_{1}}$ in Eq.~(\ref{eq:1}) by replacing
$v\to v+i\frac{D}{M}$ (reparameterisation invariance
\cite{Luke:1992cs} constrains this form), we also need to consider the
term
\begin{equation}
-2\frac{J_{qN}}{M}i D^{\beta }\left\{ \overline{N}\left[ S^{\mu
_{0}},S_{\beta }\right] \left(v+i\frac{D}{M}\right)^{\mu _{1}}N\left[
1+\eta N^{\dagger }N\right] 
\right\} \,,  \label{Oq}
\end{equation}
where $J_{q(g)N}$ is the quark(gluon) angular momentum content of the
nucleon, to obtain 
\begin{equation}
J_{qA}=\langle x\rangle _{qN}L_{z}+2J_{qN}S_{z}\left[ 1+\eta \,\mathcal{H}%
(A) \right] \,.
\end{equation}
Remarkably, explicit calculation \cite{futurework} shows that the $L_{z}$
term is free from two-body current corrections even though $\alpha _{1}\neq
0 $ in Eq.~(\ref{eq:1}). Similarly, the gluon contribution to nuclear spin
satisfies 
\begin{equation}
J_{gA}=\langle x\rangle _{gN}L_{z}+2J_{gN}S_{z}\left[ 1-\eta \,\mathcal{H}%
(A) \right] \,,
\end{equation}
where the same constant $\eta $ appears by total angular momentum
conservation. Consequently, using the sum rules $\langle x\rangle
_{qN}+\langle x\rangle _{gN}=1$ and $J_{qN}+J_{gN}=1/2$, we recover 
\begin{equation}
J_{qA}+J_{gA}=L_{z}+S_{z}\,.
\end{equation}
For details, see \cite{futurework}.

To summarise, we have studied the EMC effect (nuclear modification of parton
distributions) in EFT and seen that the scale separation of short and long
distance effects provides a model independent derivation of the
factorisation of the $x$ and $A$ dependencies of $R_{A}(x)$, relying only on
the symmetries of QCD. Similar factorisations are predicted to occur in
other probes of nuclear structure such as spin-dependent structure functions
and GPDs.

\textit{Acknowledgements:} We thank P. Bedaque, E.~M.~Henley, D. Lee, W.
Melnitchouk, G.~A.~Miller, A.~S.~Rinat, M. J. Savage, A.~W.~Thomas and
W.-K.~Tung for useful comments and discussions. This work was supported by
the US Department of Energy under contract DE-FG03-97ER41014 and the
National Science Council of ROC.

\end{document}